\documentclass[fleqn,10pt]{SelfArx} 

\usepackage{lipsum} 
\usepackage{psfig}
\usepackage{url}
\usepackage{color}

\def\etal{{\rm et~al.\ }}

\def\kms{{\rm \;km\;s^{-1}}}

\def\kmsmpc{\kms\;{\rm Mpc}^{-1}}

\def\simlt{\lower.5ex\hbox{$\; \buildrel < \over \sim \;$}}
\def\simgt{\lower.5ex\hbox{$\; \buildrel > \over \sim \;$}}

\definecolor{color1}{RGB}{0,0,90} 
\definecolor{color2}{RGB}{0,20,20} 

\definecolor{color1}{RGB}{0,0,90} 
\definecolor{color2}{RGB}{0,20,20} 

\usepackage{hyperref} 
\hypersetup{hidelinks,colorlinks,breaklinks=true,urlcolor=color2,citecolor=color1,linkcolor=color1,bookmarksopen=false,pdftitle={Title},pdfauthor={Author}}                                                                  
                                                                                                                                                                                                                             
\JournalInfo{Quarterly Physics Review  No 1 (2015) pp 1-14} 
\Archive{} 
                                                                                                                                                                                                                             
                                                                               \PaperTitle{On the measurement of cosmological parameters} 

\Authors{Rupert A.C. Croft\textsuperscript{1,2}*, Matthew Dailey\textsuperscript{1,2}} 
\affiliation{\textsuperscript{1}\textit{McWilliams Center for Cosmology,
Carnegie   Mellon  University, Pittsburgh, PA 15213, USA}} 
\affiliation{\textsuperscript{2}\textit{Dept.   of  Physics,   Carnegie   Mellon  University,
Pittsburgh, PA 15213, USA}} 
\affiliation{*\textbf{Corresponding author}: rcroft@cmu.edu} 
                                                                                                                                                                                                                             
\newcommand{\keywordname}{Keywords} 

\Abstract{We have catalogued and analysed 
cosmological parameter determinations 
and their error bars published
between the years 1990 and 2010. Our study focuses on the 
popularity of measurements, their precision and their accuracy.
The accuracy of past measurements is gauged by comparison
with the WMAP results of Komatsu et al. (2011).
The 637 measurements in our study are of 12 different
parameters and we place the techniques used to carry them out into 12
different categories.
We find that the popularity of parameter measurements (published
measurements per year) in all 12 cases except
for the dark energy equation of  state parameter $w_0$ peaked between
1995 and 2004. Of the individual techniques, only Baryon Oscillation
measurements were still rising in popularity at the end of the studied
time period. The quoted precision (fractional error)
of most measurements has been declining
relatively slowly, with several parameters, such as the amplitude
of mass fluctutations $\sigma_{8}$ and the Hubble constant $H_{0}$
remaining close to the $10\%$ precision level  for a 10-15 year
period.
The accuracy of recent parameter measurements is generally what would be
expected given the quoted error bars, although before the year 2000,
the accuracy was significantly worse, consistent with an 
average underestimate of the error bars by a factor of $\sim 2$.
When used as complement to traditional forecasting techniques, our
results suggest that future measurements of parameters such as fNL,
and $w_{a}$ will have been informed by the gradual improvment in understanding
and treatment of systematic errors and are likely to be accurate. However,
care must be taken to avoid the effects of confirmation bias, which may
be affecting recent measurements of dark energy parameters.
For example, of the 28 measurements of $\Omega_{\Lambda}$ in our sample
published since 2003, only 2 are more than 1 $\sigma$ from the WMAP
results. Wider use of blind analyses in cosmology could help to avoid this.}

\begin{document}

\flushbottom 

\maketitle 

\tableofcontents 



\section{Introduction}
Modern cosmological parameters have been 
measured since Hubble's (1929) discovery of 
the expansion of the Universe. The number of model parameters
increased during the late 1980s with the introduction of what is 
often referred to as the ``Standard cosmological model'' (e.g.
Dodelson 2005). The
idea of ``Precision
cosmology'' emerged more recently, and by the present time,  many of 
the parameters in this model are well known (see e.g., 
Komatsu \etal 2011, hereafter WMAP7). This
presents us with an interesting opportunity: by comparing the past measurements
of parameters and their error bars with the currently known values, we 
can evaluate how well the measurements were carried out in the past,
how realistic the quoted uncertainties were, and which methods gave the 
most statistically reliable results. We can also study 
how both their precision and accuracy has varied with time. Such research
will help us in our quest to make critical evaluations of what will
be possible in the future, and by working with past data
serves as a complement to more conventional future extrapolations 
of technology and techniques (e.g., the report of the
Dark Energy Task Force, hereafter DETF,
Albrecht \etal, 2006). In the present paper we make a 
first attempt at such a study, by compiling published  parameter 
values taken from the NASA Astrophysics Data System over the years 1990-2010.

Previous studies of cosmological parameter determinations have tended to
focus on the Hubble Constant, $H_{0}$, for which there is a
longer than 80 year baseline for analysis. Several
papers have used  the comprehensive
database compiled by 
John Huchra\footnote{https://www.cfa.harvard.edu/~dfabricant/huchra/}
 to generate their dataset, such as  
 the study of the  
non-Gaussian error distribution in those measurements (Chen \etal 2003).
Gott \etal (2001)  used median statistics in a metanalysis of these
$H_{0}$  measurements to find the most probable value (and also analysed
early measurements of $\Omega_{\Lambda}$).
This median statistics approach has also been used to combine individual
estimates of $\Omega_{m0}$, the present mean mass density in non-relativistic
matter
by Chen and Ratra (2003). In the present paper we do not seek to
combine the measurements from various works into best determinations
of parameters. Instead we start from the assumption that the parameters
we look at have been well measured (and their correct values are close to
the
WMAP7 values) and see what this implies about past measurements.

We therefore will be starting with the assumption that $\Lambda$CDM is
the correct cosmological model. This should be borne in mind when interpreting
our results. Even if the true cosmology turns out in the future to be
something else, we expect that the effective values of the $\Lambda$CDM
parameters are not likely to be very different (given the good fits to current
data), so that our approach will have some value even then. Parameters
which at the moment are unknown, or very poorly constrained, such as
the non-Gaussianity parameter fNL (e.g., Slosar \etal 2008),
or the time derivatives
of the dark energy equation of state parameter $w$ (e.g.,
Chevallier \& Polarski 2001) can
obviously not be studied at present with our approach. Instead we
hope that the general lessons from the past about the reliability of 
error bars, methods and achievable precision and accuracy can usefully to
inform future efforts to measure those parameters.

The DETF report explains how four different techniques are being used and
will be used in the future to constrain dark energy parameters. These
techniques, gravitational lensing, baryon oscillations, 
galaxy cluster surveys and supernova surveys all have a history and have
been involved in a large number of previous measurements of different
parameters.
It is interesting to see how they have performed in the past, and
evaluate  them based on this data. By looking over the published
record, we can also show how measurement precision has changed, in 
terms of the quoted fractional error bars, and see how this compares
with predicted future trends. One can ask whether for example the
earlier error bars were unrealistically small, so that the quoted precision
of measurements has not changed much. This should have consquences for
the accuracy of measurements, which we will define and measure.
In general, our motivation for this study can be summarized by the idea
that once cosmological parameter measurements are published, for the most 
part they are ignored when future work arrives. The dataset left behind
can instead become a valuable resource to inform future work.

Our plan for the paper is as follows. In Section 2., we detail 
the source for the cosmogical parameter estimates and how the data was
collated. We explain the different categorizations of measurements 
and methods and standardization that was carried out. In Section 3 we outline
the steps involved in our analysis of the data, and present
results including historical
trends in some individual parameters and the precision and accuracy of
measurements. In Section 4 we summarize our findings
and discuss our results.

\section{Data}

We have made use of the NASA Astrophysics Data System
\footnote{http://adsabs.harvard.edu/} to generate our dataset by carrying
out an automated search of publication abstracts for the years (1990-2010). We
limited the search to published papers which include cosmological
parameter values and their error bars in the paper abstract itself. It is
of course possible to carry out a more extensive analysis by searching the
main text of each paper, and we estimate from a random sampling that 
approximately $40\%$ of parameter estimates are missed by our abstract-only
technique. We make the assumption that this does not bias our sample.
The total number of parameter measurements in the 20 year period shown
is 637.

\subsection{Parameters}

The search  we use in the ADS abstract query form is a search for
the following terms: ``sigma8'',''H0'', ``Omega'',''Lambda'',
''m$\_$nu",''baryon''. We also restrict our search to the following
journals: MNRAS, Astrophysical Journal, ApJ Letters, ApJ supplement,
and Physical Review Letters. This parameter
search query appears restrictive, but 
enables results for 12 different parameters to be found, including 
associated parameters. These 12 are:
\begin{enumerate}
\item $\Omega_{M}$, the ratio of the present matter density to the critical density.
\item $\Omega_{\Lambda}$, the cosmological constant as a fraction of the
  critical density,
\item $H_{0}$, the Hubble constant,
\item $\sigma_{8}$, the amplitude of mass fluctuations,
\item $\Omega_{b}$, the baryon density as a fraction of the
  critical density,
\item  $n$, the primordial spectral index
\item $\beta$, equivalent to $\Omega_{m}^{0.6}/b$ where $b$ is the galaxy bias,
\item $m_{\nu}$, the neutrino mass,
\item $\Gamma$, equivalent to $\Omega_{m}H_{0}$/100 $\kmsmpc$,
\item $\Omega_{m}^{0.6}\sigma_{8}$, a combination that arises in peculiar
velocity and lensing measurements,
\item $\Omega_{k}$ the curvature,
\item $w_{0}$, the equation of state parameter for Dark Energy. 
\end{enumerate}

The measurements are generally quoted with 1 $\sigma$ errors on the parameters
but $7\%$ have 2 $\sigma$ errors. In this case, in order to have a uniform
sample, we halve the 2 $\sigma$ error bars. We have tested the effect of
ignoring excluding these $7\%$ of measurements on our results (Section
3.3) and find
that our conclusions are insensitive to this. Some of the measurements are
also quoted with separate systematic and statistical error bars (6$\%$ of
the sample). In this case we sum the statistical and systematic errors
to make a total error bar. We also test the effect of
 adding them in quadrature,
or ignoring the systematic part altogether (see Section 3.4).

Given that our approach is to assume that the WMAP7 results are correct within
their quoted errors
and that the $\Lambda CDM$ model describes the observations
well, we use the $\Lambda CDM$ model
values  to convert combinations
of published parameters into those listed above. For example, when
measurements
of $\Omega_{b}h^{2}$ are given we convert these into a value for 
$\Omega_{b}$ using the WMAP7 value of $h=0.702$. For  reference we give
our fiducial values of each parameter in Table \ref{fidparams}. 
As stated in the caption, most of these are
taken from  Table 1 in WMAP7,  but others are assumptions based on 
$\Lambda CDM$ (e.g. $w_{0}=-1$ exactly).

\begin{table}[t]
{
\caption[fidparams]{\label{fidparams}
Fiducial values for the cosmological parameters used in this paper.
These values are used when computing the accuracy of past measurements.
All parameters are taken from the last column of Table 1. in 
the WMAP7 paper, which are mean of the 
posterior distribution of combined WMAP+BAO+$H_{0}$ measurements (we
have also tried the maximum likelihood parameters, with no difference in 
our results),
except parameters (ii),(viii),(xi) and (xii) for which we have
assumed that an exactly flat $\Lambda$CDM model holds with $m_{\nu}=0\pm0.1
eV$. The quoted error bars are derived from the WMAP7 error bars, with the
exception of parameter (vii) for which an error bar of 0.1 is used to
approximately take into account differences in galaxy bias between
different samples. We explore the effect of adding these
error bars in quadrature to the error bars of past measurements 
in Section 3.4}

\begin{tabular}{ccc}
\hline&\\
Parameter  & Central value &1 $\sigma$ error bar  \\
\hline & \\
(i)  $\Omega_{M}$ & 0.274 & 0.013 \\
(ii) $\Omega_{\Lambda}$ & 1.0-0.274 & 0.013 \\
(iii)  $H_{0}$ & 70.2 kms$^{-1}$Mpc$^{-1}$ & 1.4 kms$^{-1}$Mpc$^{-1}$ \\
(iv) $\sigma_{8}$ & 0.816 & 0.024 \\
(v) $\Omega_{b}$ & 0.0458 & 0.0016\\
(vi) $n$ & 0.968 & 0.012 \\ 
(vii)  $\beta$ & 0.460 & 0.1 \\
(viii) $m_{\nu}$ & 0.0 eV& 0.1 eV\\
(ix)  $\Gamma$ & 0.193 & 0.006 \\
(x) $\Omega_{m}^{0.6}\sigma_{8}$ & 0.376 & 0.015 \\
(xi) $\Omega_{k}$ & 0.0 & 0.0 \\
(xii)  $w_{0}$ & -1.0 & 0.0 \\
\hline&\\
\end{tabular}
}
\end{table}

\subsection{Measurement Methods}

For each published measurement, we also choose a category based on the
type of data and method used to extract the cosmological parameter. There 
are obviously many different possible choices of categorization possible
and with different coarseness. We choose the following 12 categories in
order to have a reasonable number of measurements in each (the mean is 53):
\begin{enumerate}
\item Cosmic Microwave Background (CMB), specifically measurement of
primary anisotropies,
\item Large-Scale Structure (LSS), which includes clustering
of galaxies, galaxy clusters (BAO measurements and redshift 
distortions are considered separately).
 the Ly$\alpha$ forest, quasar absorption lines, and quasars.
\item Peculiar velocities, which includes measurements of galaxy 
peculiar velocities inferred from distance measurements and 
redshifts, and the cosmic dipole,
\item Supernovae, which includes techniques that
use supernova distance measurements.
\item Lensing, which includes constraints from the number of strong
gravitational lenses, weak lensing shear, and  gravitational lens time delay, 
\item Big Bang Nucleosynthesis (BBN),
\item Clusters of galaxies including their abundance and their masses. Includes 
Sunyaev-Zeldovich measurements,
\item Baryonic Acoustic Oscillation measurements from large-scale structure
of galaxies and clusters,
\item The Integrated Sachs Wolfe effect (ISW),
\item $z$ distortions, redshift distortions of clustering
\item Other, includes Tully Fisher distance estimates, galaxy ages and/or
colours, globular cluster distances, internal structure of galaxies, cepheid
distances,  surface brightness fluctuations, reverberation mapping,
radio source size, and Gamma Ray Burst distances,
\item Combined, includes measurements that result from a combination
of techniques or past measurements, without the addition of new measurements.
\end{enumerate}

\begin{figure}
\centerline{
\psfig{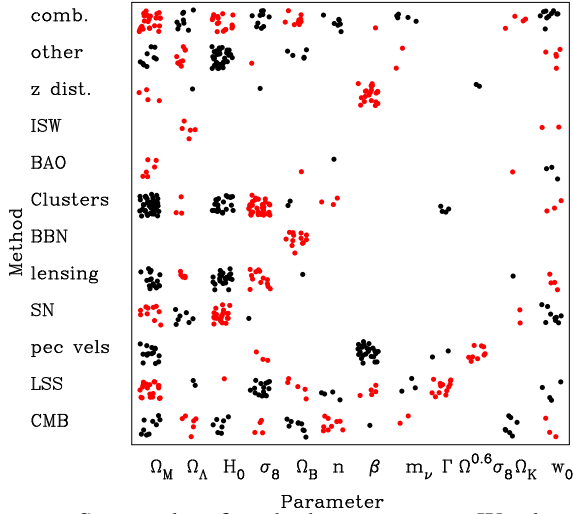}
}
\caption{
Scatter plot of method vs. parameter. We plot as a point each of
the 637 published measurements, with $y$-axis representing the 12 method
bins of Section 2.2 and the $x$-axis the 12 parameters of Section 2.1.
In order to make the points visible we have added a random offset of a 
fraction of the bin width to each point. The red and black colours
are used solely to enhance differentiation between the bins. 
\label{methodvsparam}
}
\end{figure}

In Figure \ref{methodvsparam} we show a scatter plot of method vs parameter
for our dataset. We can see that the most popular parameter/method combination
is $\Omega_{M}$ measured using galaxy clusters, but that in general there
is a fairly wide selection of method and parameter, with just over half
(76 out of 144) of the combinations covered by at least one published abstract.

\section{Analysis}
Our analysis is in two parts, the first being a study of general trends
in the number of parameter measurements and popularity of different methods
by year, as well as a looking at the measurement value vs year for
a subset of parameters (Sections 3.1 and 3.2). In the second part
(Sections 3.3 and 3.4), we 
compute the precision and accuracy of the measurements and see how these
have varied with time.

\begin{figure}
\centerline{
\psfig{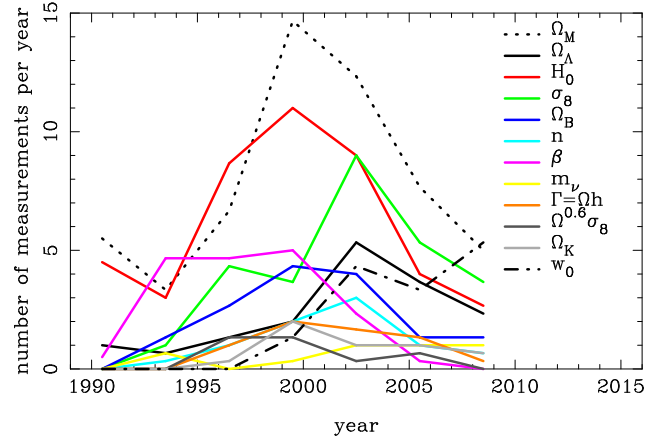}
}
\caption{
Number of cosmological parameter measurements published per year, with curves
representing the
12 different parameters listed in Section 2.1. Bins of width 3 years
were used to compute the curves.
\label{nvsyearparam}
}
\end{figure}

\subsection{Number of studies by year}
In Figure \ref{nvsyearparam} we show how the number of parameter measurements
per year has varied with time. The results are shown averaged in bins of 3
years.

It is immediately noticeable that nearly all of the
parameters have a peak in the number of measurements
around the years 2000-2003, and then a decline in the post-WMAP1
(Spergel \etal 2003) era. Exceptions to this are measurements
of $w_{0}$, which are still increasing in number, and constrains
on $m_{\nu}$. Of course this historical trend is largely guaranteed
by our selection of the parameter set we have chosen, which in
large part are considered to have been well measured already. Other 
parameters such as fNL, w$_{a}$, or the modified gravity parameter
$E_{G}$ (see e.g., Reyes \etal 2011) would still be increasing on such a plot.
Figure \ref{nvsyearparam} can also be viewed as a measure of the extent
to which parameters are considered to be well measured. For individual
parameters such as $\sigma_{8}$, there are still many measurements published
even at the current time, but the decline is still there.

\begin{figure}
\centerline{
\psfig{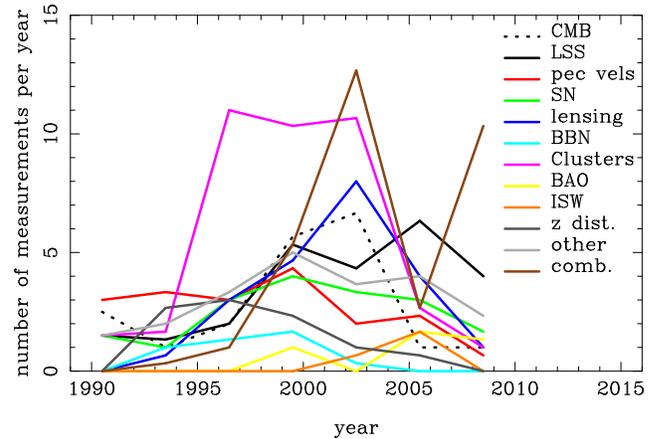}
}
\caption{
Number of cosmological parameter measurements published per year, with curves
representing the
13 different measurement methods listed in Section 2.1. Bins of width 3 years
were used to compute the curves.
\label{nvsyear}
}
\end{figure}

Another way to present the data is shown in Figure \ref{nvsyear},
where the popularity of different methods with time can be examined.
Here it can be seen that ``combined'' methods are the exception
to the general post WMAP1 decline. In overal number, galaxy clusters have
proven the most popular cosmological probe, with a sharp  start in the 
early 1990s. Supernovae and Large-Scale structure measurements have
remained fairly constant since 2000, and the popularity of gravitational 
lensing per year has not been much different from that of galaxy clusters,
except lagging behind by about 8 years. BAO measurements are the only
technique still on the rise, reflecting the 
current and future large-scale structure surveys targeted at BAO (e.g.,
Eisenstein \etal 2011, Blake \etal 2011, Schlegel \etal 2011).

\begin{figure}
\centerline{
\psfig{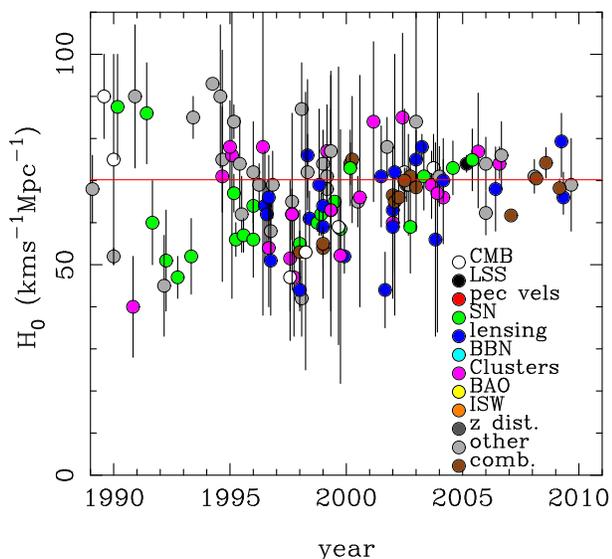}
}
\caption{
Individual published values of the Hubble constant, $H_{0}$ as a function
of year. We show one sigma error bars on the points, and the point colour 
(shown in the legend)
denotes the technique used to make the measurement (see Section 2.2 for
more details).
\label{h0}
}
\end{figure}

\subsection{History of individual parameter measurements}
\begin{figure}
\centerline{
\psfig{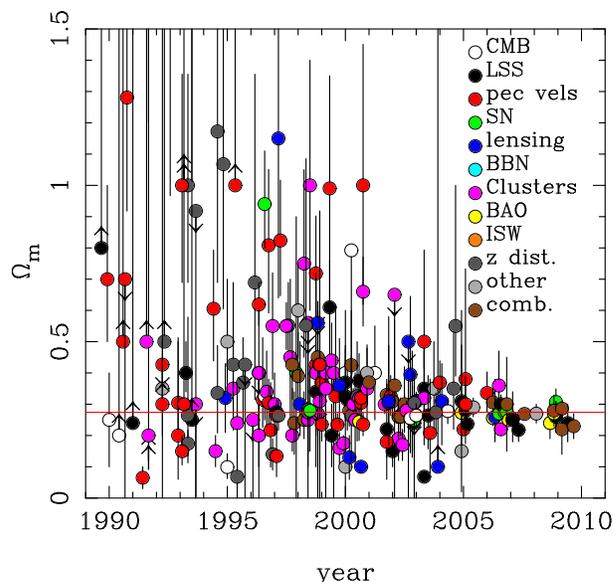}
}
\caption{
Individual published values of the density
parameter, $\Omega_{\rm m}$ as a function
of year. We show one sigma error bars on the points, and the point colour 
(shown in the legend)
denotes the technique used to make the measurement (see section 2.2 for
more details). 1 (2) sigma upper and lower limits are shown using
single (double) arrows.
\label{omega}
}
\end{figure}

\begin{figure}
\centerline{
\psfig{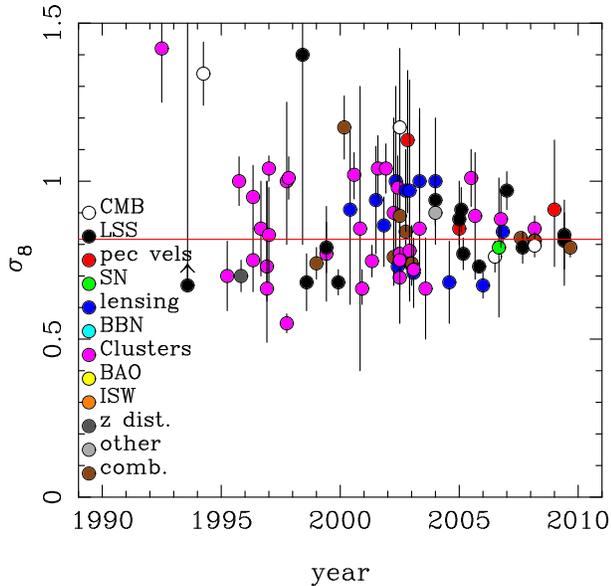}
}
\caption{
Individual published values of the amplitude of mass fluctuations,
 $\sigma_{8}$ as a function
of year. We show one sigma error bars on the points, and the point colour 
(shown in the legend)
denotes the technique used to make the measurement (see section 2.2 for
more details). 
\label{sigma8}
}
\end{figure}

\begin{figure}
\centerline{
\psfig{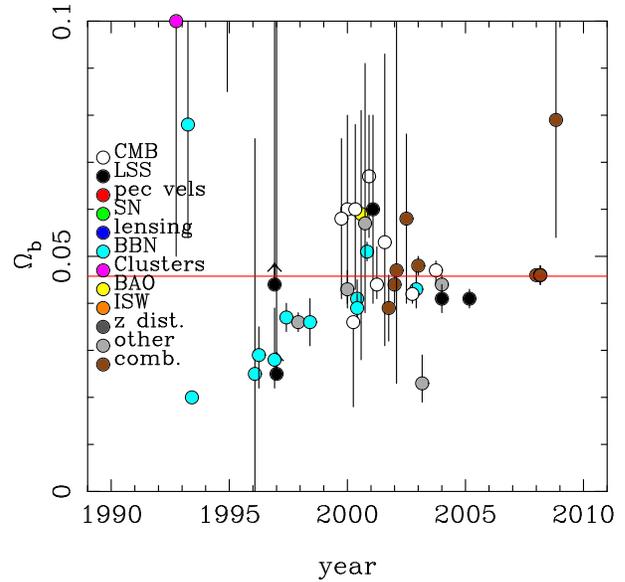}
}
\caption{
Individual published values of the baryon density as a fraction
of the critical density, $\Omega_{\rm b}$, as a function
of year. We show one sigma error bars on the points, and the point colour 
(shown in the legend)
denotes the technique used to make the measurement (see section 2.2 for
more details). 1 (2) sigma upper and lower limits are shown using
single (double) arrows.
\label{omegab}
}
\end{figure}

\begin{figure}
\centerline{
\psfig{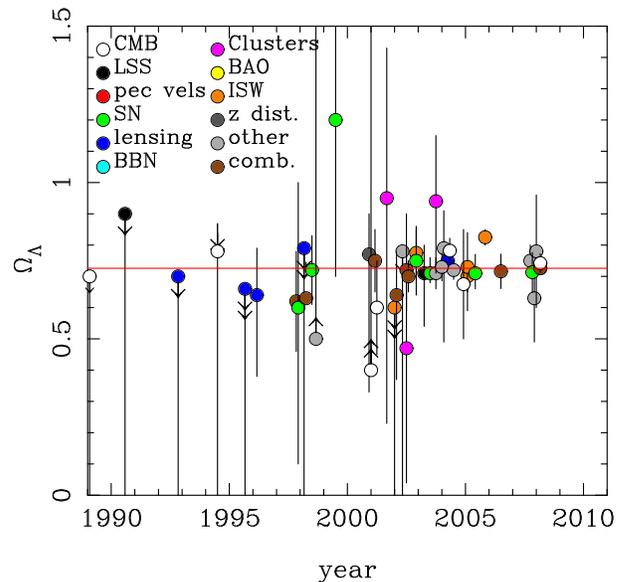}
}
\caption{
Individual published values of the vacuum density parameter
 $\Omega_{\Lambda}$ as a function
of year. We show one sigma error bars on the points, and the point colour 
(shown in the legend)
denotes the technique used to make the measurement (see Section 2.2 for
more details). 1 (2) sigma upper and lower limits are shown using
single (double) arrows.
\label{lambda}
}
\end{figure}

\begin{figure}
\centerline{
\psfig{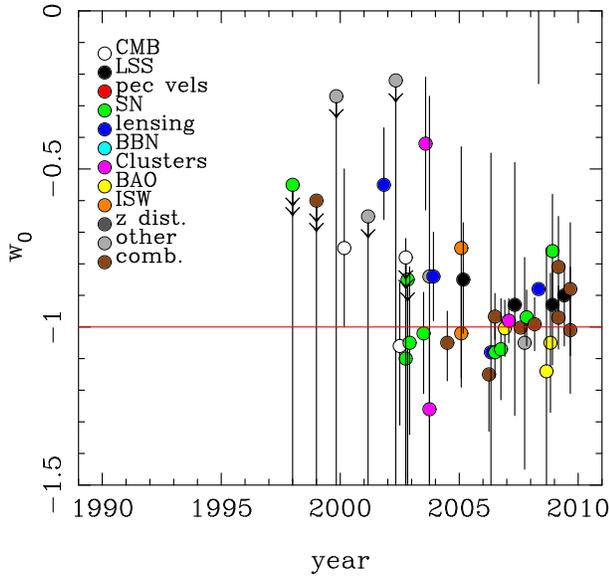}
}
\caption{
Individual published values of the dark energy
equation of state parameter $w_{0}$ as a function
of year. We show one sigma error bars on the points, and the point colour 
(shown in the legend)
denotes the technique used to make the measurement (see section 2.2 for
more details). 1 (2) sigma upper and lower limits are shown using
single (double) arrows.
\label{w0}
}
\end{figure}

It is instructive to study the distribution of data points and their
error bars as a function of time, and we do this for a subset of parameters in
Figures \ref{h0} through \ref{w0}. In each case we show the WMAP7 best 
fit value for the parameter as a horizontal line. This type of plot
is most familiar from the studies of Huchra for
the Hubble constant, where the initial values reported by Hubble were over 5 
times the currently accepted values.

In Figure \ref{h0} we show how Hubble constant determinations have changed
over the last 20 years, with the beginning of this time period
overlapping with the end of the  $\sim20$  year timeframe during which 
measurements of $H_{0}$ were largely divided into two groups, one group
closer to $50 \kmsmpc$ (e.g., Sandage \& Tammann  1975), and one closer to 
$100 \kmsmpc$ (e.g., Devaucouleurs \etal 1979). These two 
camps can be seen prior
to 1995 in  Figure \ref{h0}, where it is also obvious that their
error bars are largely not compatible, or indeed compatible with the eventual
currently favoured value of $H_{0}=70 \kmsmpc$.
 The HST Key Project (hereafter KP)
to measure
the extragalactic distance scale published its first results in 1994
(Freedman \etal 1994), and final results in 2001 (Freedman \etal 2001).
The main contribution of the project was to extend the Cepheid-based rung
of the distance ladder to cosmological distances. Freedman \etal 2001
combined this with other datasets (Type IA and type II 
SN, the galaxy Tully-Fisher relation,
surface brightness fluctuations and galaxy fundamental plane) to yield 
different measurements which were all consistent with $H_{0}=72\pm8 \kmsmpc$,
meeting the goal of a $\sim10\%$ measurement of $H_{0}$. 

This post-1994 period of activity related to the KP is immediately apparent in 
Figure \ref{h0}. It can also be seen that different methods have produced
results which were somewhat divergent at first but which eventually became
consistent with the final result by the end of the 1994-2001 KP period.
An example of this is the determination from type IA supernovae,
where it can be seen that the green points representing these track
steadily upwards from 1993 onwards. A large cluster of gravitiational lensing 
time delay measurements also exhibits a similar trend, and indeed some
other measures such as galaxy cluster Sunyaev-Zeldovich measurements
are somewhat lower than $H_{0}=70\kmsmpc$. This set of lower results
largely disappears by 2003, which is when the next sudden tightening of
determinations occurs, concident with the WMAP1 data release. The WMAP1
best fit value of  $H_{0}$ was $72\pm5\kmsmpc$, and after that date
essentially
all measurements are consistent with it. Of course the WMAP1 result was 
strikingly similar to the KP result even though it involved radically
different physics. The evidence of Figure \ref{h0} is that the combination
of the two sets of measurements was enough to convince most researchers that
the measurement goal had been reached. In the future, a measurement
of $H_{0}$ to even higher accuracy will be needed to make truly
accurate constraints on dark energy parameters (see e.g. the DETF
report).

There are a few obviously discordant points, for example, Leith \etal (2008)
find $H_{0}=61.7^{+1.2}_{-1.1}\kmsmpc$ from a combined analysis of several
datasets. Their analysis is not in the context of the LCDM model, but in
one in which cosmological averaging can be used to understand the acceleration
of the Universe (Wiltshire 2007).

In Figure \ref{omega} we show the history of measurements of $\Omega_{\rm m}$,
the most frequently measured parameter in our dataset. In this case we can
see that before 1999 approximately 1/3 of the measurements were consistent
with high values of $\Omega_{\rm m}\sim 0.5-1.0$, and that the most popular
technique in this early period involved the use of galaxy peculiar velocities.
The error bars were large, although there are several points which are not
consistent with the eventual WMAP7 value of  $\Omega_{\rm m}=0.274\pm0.013$.
After 1999, although peculiar velocities continued to be popular, the 
measurements were no longer sampling the high $\Omega_{\rm m}$ end
of parameter space. As with the $H_{0}$ results a second significant
tightening of published values around the final range took place in
the years 2004-2005, shortly after the WMAP1 results.

The amplitude of mass fluctuations, $\sigma_{8}$ is examined in 
Figure \ref{sigma8}. In this case we can see that the abundance
of galaxy clusters is easily the most popular method used to
measure this parameter, and the effort started in earnest
around 1995. The cluster measurements of  $\sigma_{8}$ are roughly
evenly spread around the WMAP7 value of  $\sigma_{8}=0.816\pm0.024$
until after the
WMAP1 release, when low values (below $\sigma_{8}\sim0.8$)
ceased to be published. As with the other 
parameters, the evidence of post WMAP1 tightening
is there. Lensing determinations of   $\sigma_{8}$ seemed to favour
high values,  $\sigma_{8}\sim 1$ until after WMAP1. 

Turning to the baryon density parameter $\Omega_{b}$
in Figure \ref{omegab}, we can see that the measurements are mainly
concentrated in an 8 year period between 1996 and 2004. Over this time
span two features can be clearly seen, the first being the steady rise
in $\Omega_{b}$ measured  using BBN, and other being the start of CMB
measurements around 2000. Because a high Deuterium to Hydrogen ratio
(easier to see) implies a low value of $\Omega_{b}$ this may account 
for the difficulty encountered in early BBN measurements. Both CMB and
BBN were consistent, however well before the WMAP tighening which occured
around 2003-2004, as with the other parameters.

In Figure \ref{lambda} we plot the measurements of $\Omega_{\Lambda}$. In
this case, many of the early points are upper limits which were just
consistent with the eventually measured value. The first Type 1A supernova 
results showing acceleration appeared at the end of this era of
upper limits.  The probably WMAP-related tightening of results around 2003
is especially pronounced in this plot, where one can see the published error
bars sizes immediately dropping. It is interesting to note that after 2002,
almost all measurements of $\Omega_{\Lambda}$ are consistent with the fiducial
value from Table 1. Of the  most recent 28 measurements shown in 
Figure \ref{lambda} (these are those that contribute to the last 2 points in
the   $\Omega_{\Lambda}$ accuracy plot, Figure \ref{accp}, in Section 3.4),
only 2 are more than 1 $\sigma$ from the ``correct'' value.
The sum of $\chi^{2}$ values when we compare to the  $\Omega_{\Lambda}$ from 
Table 1 per data point is 22.7 for these 28 measurements,
 which does not sound very small. However, this includes the measurement of 
Cabre \etal, (2006), which is 4.0 $\sigma$ from the Table 1. value. Without
this outlier, the  $\chi^{2}$ per data point is only 0.26.
 This could be
a signature of overestimation of the error bar size, or perhaps 
of ``confirmation bias''. We will return to this in Section 4.

The final parameter for which we examine the individual measurements
is the dark energy equation of state parameter $w_{0}$, which we
show in Figure \ref{w0}. In this case there are no measurements or
limits before the SN measurements of the acceleration of the
Universe in the late 1990s (Perlmutter \etal 1999,
Riess \etal 1998). At around the time of WMAP1 the first measurements
rather than limits on $w_{0}$ started to be published, and since then
SN have continued to be the most popular probe of this parameter. A trend
more apparent in this more recently measured parameter is the large 
number of points from ``combined'' measurements. Although one could argue that
$w_{0}$ is not at present as well known as some of the other parameters,
we have plotted the fiducial value on this graph as $w_{0}=1$ exactly. All 
measurements since 2004 (bar 2)  are consistent with this value
at the 1 $\sigma$ level.

\subsection{Precision}

One of the common themes which has emerged in the past few years
is that we are now in the era of ``precision cosmology''. It is
instructive to study what the data reveals about how we reached this point 
and what precision is currently achievable for the different parameters
and using the different techniques. We quantify the precision of measurements
to be the size of the $1\sigma$ error 
bar as a percentage of the fiducial (WMAP7)
value for each parameter. We have also tried using the error bar size
as a percentage of the quoted central value of each measurement, finding
no significant difference in our results (except
for the case of $\Omega_{\Lambda}$, for which the latter is
not a useful way of examining earlier data). In the case of the
neutrino mass, $m_{\nu}$ for which only upper limits are available, we
have taken the precision to be the limit in  $m_{\nu}$ divided by the
value of  $m_{\nu}$  required for the closure density (i.e. $\Omega_{m}=1$)
in  neutrinos, which is  $m_{\nu}=93h^{2}$eV. For $\Omega_{k}$ we divide
the measurement error by $1.0$ as the fiducial value of  $\Omega_{k}=0$.
Our definition of precision in this case is therefore more directly
related to
the precision on the measured distance to the surface of last scattering
(see e.g., Hu \& Dodelson 2002)

\begin{figure}
\centerline{
\psfig{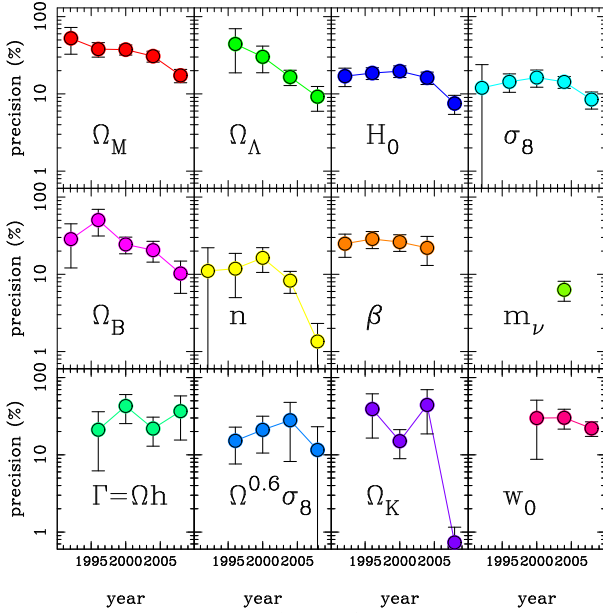}
}
\caption{
The quoted precision of measurements as a function of year for our
different cosmological parameters. The precision is defined to be the
size of the $1 \sigma$ error bar as a percentage of the fiducial parameter
value in Table \ref{fidparams}. Error bars are Poisson errors computed
from the number of measurements in each bin.
\label{prepa}
}
\end{figure}

In Figure \ref{prepa} we show the average  precision for each parameter
as a function of year. We have binned the measurements into bins of width 4
years, and when computing the average precision compute an
unweighted mean from the
measurements in each bin. We show Poisson error bars on the mean
precision.  Apart from  $m_{\nu}$, we have not included any upper or
lower limits on parameters in this plot, only published measurements of
values with error bars.

It is apparent from the general appearance of Figure \ref{prepa} that 
the precision of most measurements has not increased very steeply. The log
scale of the $y$-axis is partly responsible for this impression,
but even so, of the 12 parameters shown, 6 have a mean
precision in the latest bin
which  is compatible (within $1 \sigma$) of that in the earliest bin.
It is possible that this situation has arisen because of greater understanding
of the role of possible systematic errors as time has gone on. The value of 
$\sigma_{8}$ is now known to better than $10\%$ for
an average measurement, for example, after a 
long period in which the precision did not improve. Currently
the most precisely
known parameters are the curvature  $\Omega_{k}$  and the primordial
spectral index $n$, which are both known to about $1\%$. A large group
of parameters are currently  known to about $10\%$ precision,
 from $\Omega_{\rm M}$ (17\%),
through  $\Omega_{\rm b}$, $\Omega_{\rm \Lambda}$,$\sigma_{8}$ and $H_{0}$
(7\%).

\begin{figure}
\centerline{
\psfig{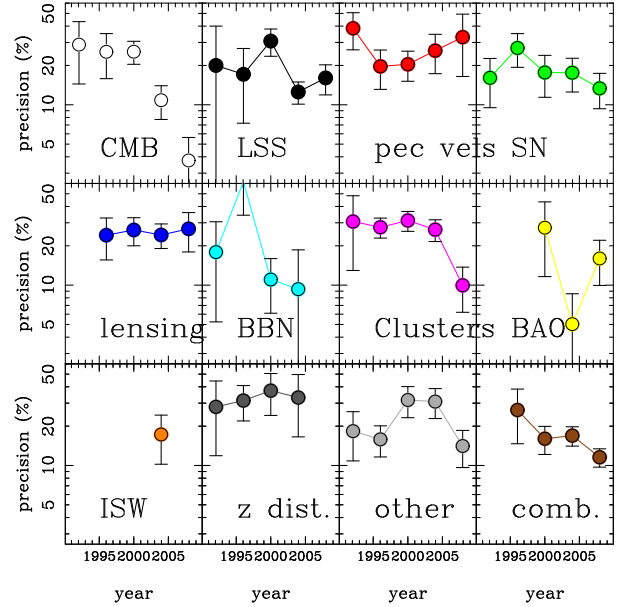}
}
\caption{
The quoted precision of measurements as a function of year for 
the different measurement techniques. The precision is defined to be the
size of the $1 \sigma$ error bar as a percentage of the fiducial parameter
value in Table 1. Each panel therefore
includes measurements made of many different 
parameters. Error bars are Poisson errors computed
from the number of measurements in each bin.
\label{precm}
}
\end{figure}

In Figure \ref{precm}, we show the precision of measurements as a function
of the technique used. As many of the techniques
are used to measure several different parameters, it is worth bearing
in mind that decreases in precision with time could be related to
the switch to a less well measured parameter. We can see that this may indeed
be happening in some cases, or else that again systematic errors are being 
confronted more as time goes in. We can differentiate between these
possibilities by considering the averaged accuracy of measurements, which
we do in the next Section. For now, we can see that lensing, 
redshift distortion and peculiar velocity
measurements have exhibited no improvement in 
quoted precision with time. The CMB on the other has improved by about an
order of magnitude over the 20 year period, and cluster measurements by about
a factor of 3. Supernova measurements are also more precise now than they were 
in the late 1990s by a factor of 2.

\subsection{Accuracy}

Our assumption that the ``correct'' values of the different cosmological
parameter values are available allows us to compute a potentially
powerful statistic, the accuracy of measurements. We define this to be the
absolute value of the difference between a measured value of a parameter
and our fidcuial value for that parameter (as listed in Table \ref{fidparams}),
divided by the quoted $1 \sigma$ error bar for that measurement. The
accuracy can therefore be written as $N_{\sigma}$, the 
average number of standard
deviations measurements are from the correct value. We note that for
a Normal distribution of errors, the average value of $N_\sigma=1$.
Values smaller than 1 indicate that the error bars have been
overestimated, and for values larger than 1 the error bars have been
underestimated. Alternatively, values smaller than 1 may also indicate
evidence for ``confirmation bias'', in which values closer to
the expected ones are favoured (not necessarily consciously).
We have chosen to use $N_{\sigma}$ as our statistic rather than the 
$\chi^{2}$  as it is more robust to outliers (not being dependent
on the square of the difference between a measurement and the true
value). Qualitatively similar conclusions would result if we did
use the $\chi^{2}$ of measurements with respect to the ``known'' values
as a measure of accuracy, however.

When computing the accuracy, one must decide how the uncertainty on 
the true values of the parameters affects the results. Two choices
which approximately bracket the range of potential effects are 
to either add the $1 \sigma$ error bars on the values in Table 1
in quadrature to the error bars on each published measurement, or else to
assume no additional uncertainty beyond the quoted error
bar for each published measurement. We have tried both, finding
almost imperceptible
quantitative differences which do not affect any of our conclusions. This can
be understood from the fact that the error bars in Table 1
are much smaller than those on past measurements.
In making our plots, we have chosen the second option, on the grounds that 
some of the uncertainty from the other measurements has already been
incorporated into the WMAP7 results we use in Table 1.

\begin{figure}
\centerline{
\psfig{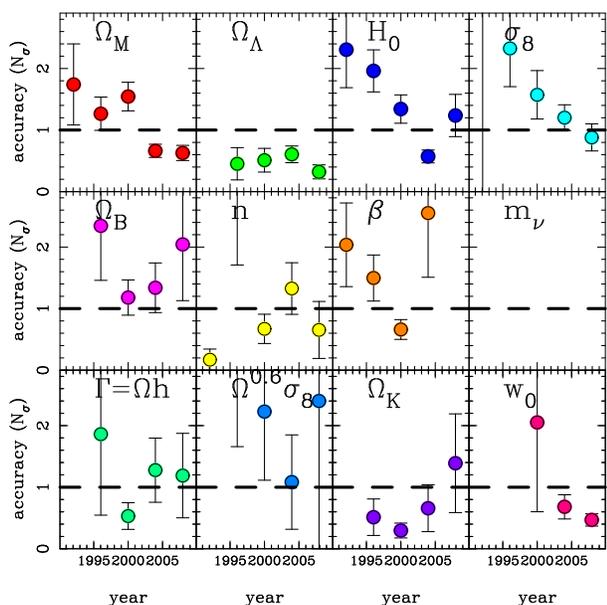}
}
\caption{
The accuracy of measurements as a function of year for 
the different parameters. The accuracy is defined to be the
difference between the quoted measurement and the 
the fiducial parameter
value in Table 1 in units of the quoted
measurement $1\sigma$ error bar. Error bars are Poisson errors computed
from the number of measurements in each bin.
The dashed line is the expectation for Gaussian statistics, $N_{\sigma}=1$.
\label{accp}
}
\end{figure}

In Figure \ref{accp} we show the accuracy for the different parameters
as a function of year, with the binning by year carried out as for
the precision (Figure \ref{prepa}), an unweighted average of the
measurements in that bin. Poisson error bars have been computed as before.
In general, one can see that the measurements in the different panels
are not extremely offset from the $N_{\sigma}=1$ line, indicating that the 
accuracy of parameter determinations has not been wildly off.  That said,
however, the $N_\sigma=1$ line is a good fit by eye in only 1 panel,
that for the shape parameter $\Gamma=\Omega h$.

Turning to individual parameters in Figure \ref{accp}, we can see
that the accuracy of measurements of $H_{0}$ and $\sigma_{8}$, has improved
over the last 20 years, so that the most recent measurements appear to have
realistic error bars. The error bars on $\Omega_{\Lambda}$ appear to
be overestimated, as do the most recent error bars on $\Omega_{M}$. From this
it would appear that signficant work 
sucessfully understanding the overall levels
of measurement uncertainty has been carried out for $H_{0}$ and $\sigma_{8}$,
but that this has not happened for some of the other parameters.
We return to this topic in Section 4.2.

\begin{figure}
\centerline{
\psfig{file=accuracybymethod.ps,angle=-90.,width=8.0truecm}
}
\caption{
The accuracy of measurements as a function of year for 
the different meaasurement techniques. The accuracy is defined to be the
difference between the quoted measurement and the 
the fiducial parameter
value in Table 1 in units of the quoted
measurement $1\sigma$ error bar. Error bars are Poisson errors computed
from the number of measurements in each bin.
The dashed line is the expectation for Gaussian statistics, $N_{\sigma}=1$.
\label{accm}
}
\end{figure}

If the varying accuracy is more tightly related to the choice of technique
than parameter, then we can expect the plot of accuracy for
different techniques (Figure \ref{accm}) to be more instructive. Here we can
see that there are indeed some techniques which have a better track record
than others. For example the use of peculiar velocities to measure
parameters has resulted in an underestimate of error bars by a factor of roughly
2 on average (there is some improvement in the most recent two points, which 
are consistent at $1\sigma$ with $N_\sigma=1$ ). The
CMB and redshift distortions have on the other hand proven 
accurate sources of measurements for 
the whole period. Galaxy clusters were sources of measurements with 
underestimated errors in the 1990s, but in the last 12 years have 
tracked  $N_\sigma=1$ very well.

The most recent 2 points for SN and
3 for BAO appear to have overestimated error bars,
signficantly so in the case of SN (by a factor of 3). SN measurements
are those which most often quote systematic error bars (which we have
added directly to the statistical errors). We have tried two other ways
of dealing with the systematic errors, either adding them in quadrature,
or ignoring them altogether. We find that with the latter most
conservative treatment, the SN results 
yield $N_{\sigma}=0.5 \pm0.07$ and $0.77\pm0.12$ for the most recent two points.
This is an improvement, indicating that the SN systematic error bars 
may well be too
conservative. It is still an underestimate, but now of similar magnitude to
the differences seen between the accuracy=1 line and some data points on the 
``other'', ``combined''  and ``LSS'' panels.
If we allow for the possibility that the Poisson error bars on our data points
in Figure \ref{accm}
are underestimates, and that there may be correlations between measurements
in different years then this may go some way to reconciling the measurements
and their hoped for accuracy. We return to this point in our discussion below
(Section 4.2).

\begin{figure}
\centerline{
\psfig{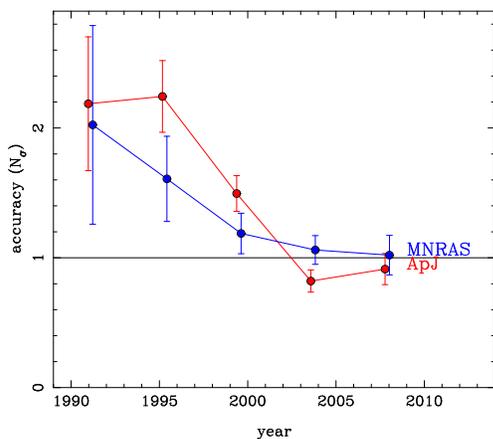}
}
\caption{
The accuracy of measurements as a function of year.
 The accuracy is defined to be the
difference between the quoted measurement and the 
the fiducial parameter
value in Table 1 in units of the quoted
measurement $1\sigma$ error bar. Error bars are Poisson errors computed
from the number of measurements in each bin. We show 
separately the accuracy for
measurements from  the two journals with the most published measurements.
\label{journal}
}
\end{figure}

One question which is not easy to answer from the multipanel Figures
\ref{accp} and \ref{accm} is how the overall accuracy of measurements is
changing by year. Are cosmological measurements improving as both
theoretical knowledge and expertise in dealing with experimental
 uncertainties improve? We can see that this does appear to be the case
by considering Figure \ref{journal}, which plots accuracy by year for
results published in the two main  journals, MNRAS and
 ApJ (including ApJL and ApJS).
These account for $35\%$ and $55\%$ of all results in our compilation,
respectively. The results before the year $\sim$2003 are significantly
inaccurate, but steadily improve with time until after this date they 
become  consistent with 
the $N_{\sigma}=1$ line. Both journals exhibit the same behaviour
within their error bars.

\begin{figure}
\centerline{
\psfig{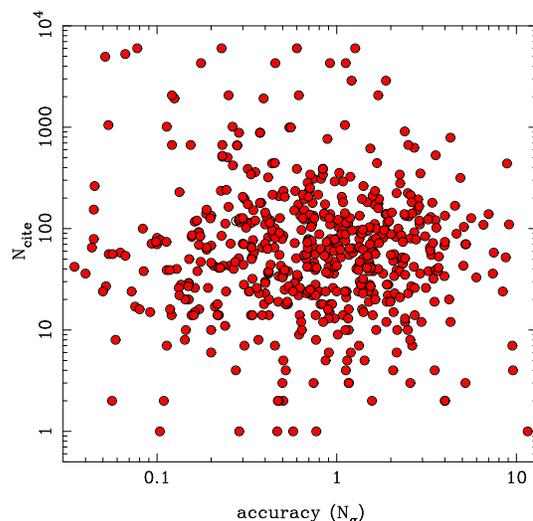}
}
\caption{
The accuracy of measurements published in a paper as a function of the
number of citations to it.
 The accuracy is defined to be the
difference between the quoted measurement and the 
the fiducial parameter
value in Table 1 in units of the quoted
measurement $1\sigma$ error bar.
\label{ncite}
}
\end{figure}

Using our tabulated data we can explore a few more aspects of the accuracy
of measurements. One can ask whether when results are published their
accuracy affects the amount by which they are cited, and therefore
whether recognition increases with the perception that measurements
are accurate. We address this in Figure \ref{ncite}, where we plot the
accuracy vs. the number of citations to a paper, both on a log scale.
We can see that there appears to be little evidence for any relationship 
between the two, so that accuracy is not an important factor in determining
the number of citations. Looking at Figure \ref{ncite}, it does seem
that there might be slightly less papers with high $N_{\sigma}$
(innaccurate) and high citations that other corners of the plot. 
This leads to a Pearson
correlation coefficient of $r=-0.066$, and therefore a slight
correlation between citations and accuracy, in that papers
with higher accuracy (lower $N_{\sigma}$) have more citations. A set of
points with no correlation would give such a result $11\%$ of the time,
so the evidence for this is marginal, however.

\begin{figure}
\centerline{
\psfig{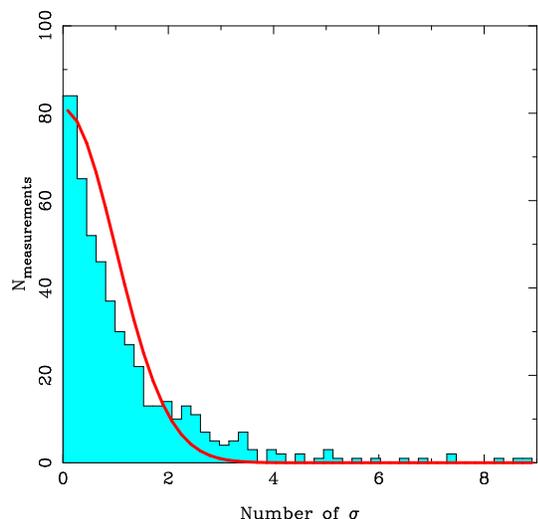}
}
\caption{
The distribution of measurement errors in units of the quoted standard 
deviation. For each measurement we divide the difference between the quoted
value and the fiducial value in Table 1 by the quoted $1 \sigma$
error bar. The results are shown as a histogram. We also show the 
expected curve for a Gaussian distribution of errors (smooth line).
\label{sigmahist}
}
\end{figure}

We note that there does exist a significant correlation between the
precision of measurements and the number of citations (not plotted).
We find a correlation coefficient of $r=-0.134$ (smaller
fractional error results are more cited) and probability $p=0.0026$
when correlating these two values. It is relatively easy
to find a possible explanation for this, as there is also a
correlation between year of  publication and precision
($r=-0.748$ , $p=9.7\times10^{-5}$),
which is just due to the overall trend in improving measurements,
 and a correlation
between year of publication and citations ($r=0.348$, $p=0.148$).
This latter is presumably due to the larger number of researchers
working in cosmology. Both of these trends combine to produce
the trend of citations with precision.

A final issue which we address when looking at the accuracy is the shape
of the error distribution. When stating that $N_{\sigma}=1$ is appropriate
for an accurate set of measurements we have made the assumption that all
quoted errors have a Gaussian distribution. This is an assumption often made
(although not by all), and is something which we can examine using our
data, by comparing the number of standard deviations that measurements
are away from our fiducial values with the curve for a Gaussian distribution.
This will tell us for example if there is a long non-Gaussian tail
to the error distribution. We show the histogram of $N_{\sigma}$ values
in Figure \ref{sigmahist} along with the Gaussian curve. The data is fairly
similar to the Gaussian curve for the low end of the  $N_{\sigma}$  range
where the majority of the data resides,
 showing that in general error bars are only slightly understimated (we have
 seen this already in Figure \ref{journal}, for example). There is however
a long tail extending to high $N_{\sigma}$ values, with some 
measurements being 8 or even 10 $\sigma$ away from their fiducial
 values. Of course with a Gaussian distribution the chance of such
events occuring would be miniuscule. We can quantify this further
by computing the fraction of measurements which are greater than
$2 \sigma$ away from the correct value. We that $19\%$ of measurements
are like this, rather than the $5\%$ expected for a Gaussian distribution.

\section{Summary and Discusssion}

\subsection{Summary}
We have compiled cosmological parameter measurements published between
1990 and 2010 and the techniques used to measure them.
Using this data we have carried out an analysis of historical trends in
popularity, precision and accuracy. The accuracy of past measurements
has been estimated by assuming that WMAP7 parameter values 
of  Komatsu \etal (2011) 
(combined with $\Lambda$CDM standard values for e.g. $w_{0}$)
are the correct ones. Our findings can be summarised as
follows:

(1) The number of published measurements for different parameters
peaks between 1995 and 2004 for all cases, except for $w_{0}$ for
which the number was still rising in 2010. 

(2) Of all techniques used to measure the parameters, only baryon
oscillation and ``combined'' measurements were still rising
in terms of publications per year by 2010.

(3) The quoted precision of measurements has been declining 
relatively slowly for most parameters, with several (e.g. $\sigma_{8}$,
$H_{0}$ remaining flat for 10-15 years.

(4) The accuracy of recent parameter measurements is generally what 
should be expected based on the quoted error bars i.e. the error
bars overall are neither understimated nor overestimated (an accuracy,
 $N_{\sigma}=1.0$, within the Poisson uncertainty on the measurement).
Before 2000, the accuracy $N_{\sigma}$ as closer to 2, indicating
underestimation of the error bars by a factor of 2. Overall, there
is a small non-Gaussian tail to the error distributions (we
find that $20\%$ of measurements are more that $2\sigma$
away from the true values.

(5) The accuracy of most methods has become consistent with $N_{\sigma}=1.0$,
with the historically most innaccurate parameter measurement
technique being the use of galaxy peculiar velocities. Measurements
of $\Omega_{M}$ and particularly $\Omega_{\Lambda}$ made since 2000 tend
to have accuracy $N_{\sigma}$ significantly less than 1.0, 
indicating ``confirmation bias'' and/or an overestimation of error bar
sizes.

\subsection{Discussion}
Over the 20 year period covered in this study, it is apparent
that many of the parameters in what is
now the concordance CDM cosmological model
went from the status of no information or only limits to being known
at the $10\%$ level or better. It is also apparent from Figure 2 that 
there was a ``golden age'' of parameter measurements between
$\sim 1995$ and $\sim 2005$ during which the number of published
measurements peaked sharply and then declined. This seems to
indicate that for many purposes (such as the use of a background
cosmology in galaxy formation models), the precision to which the
$\Lambda$CDM parameters were known by the time of the first WMAP
results is sufficient, and many of the reasons for
pinning down the model better had diminished after that.

This said, however, the exception to this rule, measurements of
$w_{0}$ (which are still rising in terms of number per year at the end
of our study) seems to point to a coming new era in parameter measurement.
Certainly, the motivation for the large number of ongoing
and future large-scale
structure, lensing and other surveys is to hunt for
the signatures of dynamical dark energy and modified gravity,
and given the number of researchers carrying out these studies
it is likely that measurements will continue to rise. Many parameters
which we have not catalogued are now within reach of
quantitiative study. These include the modified gravity
parameter, $E_{G}$ (Reyes \etal 2010)
and the time derivative of the equation
of state parameter, $w_{a}$. Measurements of such parameters 
involve searching for deviations from the concordance 
CDM model and fall into a different category from most of the parameters
we have studied in this paper.  Inflationary
parameters such as the non-Gaussianity fNL, or tensor to scalar
ratio $r$ will pinned down with higher precision in the future, and
these should also represent a growth area. The motivation for 
most future measurements being largely framed in terms of a quest for
fundamental physics, it would be logical to assume that they will continue
until the cause for the Universe's acceleration are better understood.
Likewise, parameters describing the dark matter particle should be added
to this category.

Possible behaviours for the 
precision of future parameter measurements can be predicted 
by looking at the past results (Figure \ref{prepa}).
There is a very wide range, but 
most parameters improve slowly, with a factor of 10 improvement in 
precision over the 20 years representing the extreme (2 out of 12 parameters).
The precision of some parameters has remained relatively flat for the 
whole period, so this is a possibility for future
so far unconstrained parameters. An argument
against this slow progress however is the fact that many new surveys
(such as of Baryon Oscillations) are targeted primarily at measures of
specific parameters, and this aggressive approach (for example
including specific precisions to be obtained at a given time in 
survey proposal documents) could lead to faster progress.

Our investigation of the accuracy of results could potentially lead
to some of the most interesting findings. We have seen that in the 
earlier half of our studied time period there is evidence that the
error bars were significantly underestimated, but that this
has changed over time. 

When discussing the accuracy, we are should be aware  that it was not
possible in our analysis to take into account several factors which
have the potential to affect our conclusions.
 For example, we do not keep track of
the priors that people have assumed in their measurements, and in many
of the later cases, this may include the WMAP results as priors.
That this is happening is likely to be responsible for
much of the post WMAP1 tightening of constraints seen in Figures \ref{h0}-
\ref{w0}. When computing the error bars on the mean accuracies
of measurements (Figs \ref{accp} and \ref{accm}) we have used Poisson
errors based on the number of measurements in each bin. This will tend to
underestimate the uncertainty on the accuracy because some of those
measurements could be using the same underlying data, or be using similar
priors, or a combination of the two. There will therefore be correlations
between the error bars so that our estimates of the accuracy will be
affected. Equivalently the chi-square of the fiducial result compared to
the data points will be incorrectly determined to be low because of the
correlations are not included.

Bearing the above points in mind, we return to the panels in Figures
\ref{accp} and \ref{accm} where the accuracy seems to be significantly
below $N_{\sigma}=1$. This is most obvious in the second panel
($\Omega_{\Lambda}$) of Figure \ref{accp}. Such as result
could be a sign that either the error bars have
been significantly overestimated, or else that researchers have been 
influenced by prior results (``confirmation bias''), or a combination
of the two. If we return to the data points which led to the last
two bins of panel two of Figure \ref{accp}, we find something especially
striking. Of those 28 measurements, only 2 are more than $1 \sigma$
from the fiducial results of Table 1. These 28 measurements were carried out
by approximately 11 separate groups 
(as determined by authorship lists) using several different techniques.

This closeness of published results to the ``correct'' ones is 
somewhat worrying for future measurements. One can interpret this as
coming partly from  error bars being overerestimated by cautious cosmologists,
for example by including possible systematic errors in the error
bars which are not actually present to such a large degree, or in a related
point  authors marginalizing over parameters which are actually better
known than was assumed. We note that including or excluding the
actually quoted systematic error bars (Section 3.4) has little effect
on this result. An additional question is why some parameters have
$N_{\sigma} <1$ and others do not (e.g., $\sigma_{8}$). The relatively
low number statistics of our whole dataset 
preclude us from making any strong statements about this issue.
If it does partly result from confirmation bias, one can also wonder
how
observers knew which value of $\Omega_{\Lambda}$ (for example) would be 
the ``correct'' one, given that our fiducial (mostly WMAP7) results
from Table 1 were published in 2011. If this bias is present, it
is probably related to the mean level for  $\Omega_{\Lambda}$ resulting
from several prior measurements. For example in Figure \ref{lambda}
and others, the value of the parameter seems to be pretty well
determined at least by 2003.

If we look at the techniques which are often associated with
dark energy measurements, SN and BAO, we can see in Figure \ref{accm}
that these two have low $N_{\sigma}$ for recent measurements. Of the
23 measurements which where included in the last bins of the
SN panel of Figure \ref{accm}, only 2 are more than
1 $\sigma$ from the fiducial 
result.
We note  that this fiducial result from Table 1 does include BAO
measurements, but not SN estimates of dark energy.
In the case of SN, however, only 4 measurements of $\Omega_{\Lambda}$
are included in these bins, and only 2 separate groups of researchers,
so that for that subset of data, statistical fluctuations may well be
responsible for the low $N_{\sigma}$ seen. If confirmation bias is present,
on the other hand, one could argue about who is confirming who- certainly
the first SN results on dark energy predate those from BAO and from
most other techniques. These sorts of questions might be addressed by
a more detailed look at the published measurements, including details
of priors, jointly used datasets and analysis techniques. Then again
small number statistics probably would not allow firm conclusions to be drawn.
These hints should instead serve as a warning that care and perhaps
concrete steps be taken to avoid any confirmation bias in the future.

\subsection{Case study: Future measurements of General Relativistic effects
measured from large-scale structure}

As we have seen from Section 3.1, the number of published 
measurements
per year has already peaked for many parameters. There are however certain
techniques, such as BAO which were still rising in popularity at the end 
of the study period. In this subsection, we speculate on the basis of our
study what could be the medium term (5-20 year) future of a newer probe
of cosmology, the measurement of General Relativistic (GR) 
effects in large-scale
structure, and in particular the gravitational redshift.

In the weak field limit, the gravitational redshift, $z_{g}$ of photons
with wavelength $\lambda$ emitted in a gravitational potential $\phi$
and observed at infinity is given by $z_{g}=\frac{\Delta\lambda}{\lambda}
\simeq \frac{\Delta \phi}{c^{2}}$.
Measurement of $z_{g}$ is one of the fundamental
tests of GR. First measured more that 50 years ago 
for the Earth's gravity in a laboratory setting (Pound \& Rebka 1959),
subsequent determinations were been made in the solar system
(Lopresto \etal 1991) and from spectral line shifts in
white dwarf stars (e.g., Greenstein \etal 1971). In cosmology,
theoretical predictions and attempts to measure the effect have a long
history. Light emitted in dense regions (such as galaxies in clusters or
superclusters) should be redshifted with respect to other galaxies in 
less dense regions.
Early studies of the redshift differences
between galaxy clusters of different masses (Nottale 1983) found no
effect, as did measurements made from brightest cluster galaxies (e.g.
Cappi 1995, Kim \& Croft 2004).

The first successful measurements of galaxy gravitational redshifts were made
by Wojtak et al (2011), by stacking the redshift profiles of 8000 galaxy
clusters from the Sloan Digital Sky survey. The measurement had a stated
precision of 36\% (it was a 2.8 $\sigma$ detection). It was followed by 
a number of other measurements with similar precision (Dominguez Romero, 2012,
Sadeh et al., 2015, Jimeno et al., 2015), in the fashion that might
be expected if the precision vs. time curve follows the slow decent seen 
in many of the parameters seen in Figure 10. 

At present a cosmological parameter relevant to quantify the deviations
from GR has not been uniformly used in the literature on gravitational
redshifts. The modified gravity parameter
$E_{G}$ (see e.g., Reyes \etal 2011) has emerged as a possible contender,
but gravitational redshift measurements are still at the level of quoting 
a detection significance. The difference in signal amplitude 
compared to GR for some popular 
alternative theories of gravity is approximately 30\% also (Wojtak et al 2011).

Theoretical predictions for GR effects have increased in number and in 
precision alongside the first detections. These range from early
work of Cappi (1995), Broadhurst et al. (2000), Kim \& Croft (2004) to
more recent studies which include a range of other GR galaxy clustering effects
of similar magnitude to the gravitational redshift, and which need to be taken 
into account at the same time (McDonald 2009, Yoo et al. 2012, Zhao et al. 
2013, Kaiser 2013, Bonvin et al. 2014).

Based on the results in Figures 10 and 11, 
the $\sim 30$\% precision quoted for published
measurements is likely to hold steady
for the next 10 years or so. Towards the end of this time period,
it is expected that major galaxy redshift surveys such as MS-DESI
and Euclid will allow measurements to rapidly reach the precision
of a few percent (Croft 2013).

\subsection{Conclusions}

In conclusion, we have seen that huge progress has been made in the 20 year 
period covered by our study. Important questions have been resolved (e.g.,
is the Universe open?, do massive neutrinos contribute substantially
to the dark matter density?), a model has been found which agrees with 
essentially all observational data  so far ($\Lambda$CDM), and
the parameters of that model have been pinned down at the
$1-10\%$ level. The first WMAP results
(e.g. as presented in Spergel \etal (2003)) form a watershed which is
easy to pick up in most plots of parameters with time, and serves
as a reminder that statistically measurable progress is not always 
gradual. Perhaps the most interesting aspect of our study, of the
accuracy of past results compared to our most recent knowledge has found
that understanding of systematic errors and uncertainties in cosmological
measurements has demonstrably improved since the early 1990s. On average,
results in the last 10 years are consistent with expectations, given their
error bars, something which should instill confidence in 
future measurements. There are some signs that recent measurements of dark 
energy parameters are closer to the ``expected'' values for $\Lambda$CDM
than statistically likely. These may be explainable by correlations
between measurements which we have not included. On the other hand this
may serve as a sign that as cosmology collaboration sizes increase carrying
out more blind analyses (as in particle physics) may be a good idea.

\section*{Acknowledgments}
We acknowledge partial support from NSF award AST 1412966.
RACC acknowledges useful discussions with Bill Holzapfel, Saul Perlmutter,
Jeff Peterson, Anze Slosar and Martin White. 
This research has made use of NASA's Astrophysics Data 
System Bibliographic Services.

{}

\end{document}